\documentclass[twocolumn, trackchanges]{aastex63}

\received{}
\revised{}
\accepted{\today}
\submitjournal{ApJ}

\shorttitle{\ion{H}{1} contents of nearby galaxies}
\shortauthors{Namiki et al.}
\graphicspath{{./}{figures/}}

\begin{document}

\title{What determines the \ion{H}{1} gas content in galaxies?:

morphological dependence of the \ion{H}{1} gas fraction across \mbox{\boldmath $M_*$}-\textit{SFR} plane}

\correspondingauthor{Shgieru V. Namiki}
\email{shigeru.namiki@nao.ac.jp}

\author[0000-0002-5693-4756]{Shgieru V. Namiki}
\affiliation{The Graduate University for Advanced Studies, SOKENDAI, 2-21-1 Osawa, Mitaka, Tokyo, 181-8588, Japan}
\affiliation{Subaru Telescope, National Astronomical Observatory of Japan, National Institutes of Natural Sciences, 650 North A'ohoku Place, Hilo, HI 96720, USA}

\author{Yusei Koyama}
\affiliation{The Graduate University for Advanced Studies, SOKENDAI, 2-21-1 Osawa, Mitaka, Tokyo, 181-8588, Japan}
\affiliation{Subaru Telescope, National Astronomical Observatory of Japan, National Institutes of Natural Sciences, 650 North A'ohoku Place, Hilo, HI 96720, USA}

\author{Shuhei, Koyama}
\affiliation{Institute of Astronomy, Graduate School of Science, University of Tokyo, 2-21-1 Osawa, Mitaka, Tokyo 181-0015, Japan}

\author{Takuji Yamashita}
\affiliation{National Astronomical Observatory of Japan, National Institutes of Natural Sciences, 2-21-1 Osawa, Mitaka, Tokyo 181-8588, Japan}

\author{Masao Hayashi}
\affiliation{National Astronomical Observatory of Japan, National Institutes of Natural Sciences, 2-21-1 Osawa, Mitaka, Tokyo 181-8588, Japan}

\author{Martha P. Haynes}
\affiliation{Cornell Center for Astrophysics and Planetary Science, Space Sciences Building, Cornell University, Ithaca, NY 14853 USA}

\author{Rhythm Shimakawa}
\affiliation{Subaru Telescope, National Astronomical Observatory of Japan, National Institutes of Natural Sciences, 650 North A'ohoku Place, Hilo, HI 96720, USA}


\author{Masato Onodera}
\affiliation{The Graduate University for Advanced Studies, SOKENDAI, 2-21-1 Osawa, Mitaka, Tokyo, 181-8588, Japan}
\affiliation{Subaru Telescope, National Astronomical Observatory of Japan, National Institutes of Natural Sciences, 650 North A'ohoku Place, Hilo, HI 96720, USA}



\begin{abstract}
We perform a stacking analysis of the \ion{H}{1} spectra from the Arecibo Legacy Fast ALFA (ALFALFA) survey for optically-selected local galaxies from the Sloan Digital Sky Survey (SDSS) to study the average gas fraction of galaxies at fixed stellar mass ($M_*$) and star formation rate (\textit{SFR}).
We first confirm that the average gas fraction strongly depends on the stellar mass and \textit{SFR} of host galaxies; massive galaxies tend to have a lower gas fraction, and actively star-forming galaxies show higher gas fraction, which is consistent with many previous studies. Then we investigate the morphological dependence of the \ion{H}{1} gas mass fraction at fixed $M_*$ and \textit{SFR} to minimize the effects of these parameters.
We use three morphological classifications based on parametric indicator (S\'{e}rsic index), non-parametric indicator (C-index), and visual inspection (smoothness from the Galaxy Zoo 2 project) on the optical image.
We find that there is no significant morphological dependence of the \ion{H}{1} gas mass fraction at fixed $M_*$ and \textit{SFR} when we use C-index. In comparison, there exists a hint of diminishment in the \ion{H}{1} gas mass fraction for ``smooth'' galaxies compared with ``non-smooth'' galaxies.
We find that the visual smoothness is sensitive to the existence of small-scale structures in a galaxy. Our result suggests that even at fixed $M_*$ and \textit{SFR}, the presence of such small-scale structures (seen in the optical image) is linked to their total \ion{H}{1} gas content.
\end{abstract}



\section{Introduction} \label{sec:intro}
Morphologies of galaxies are known to be strongly correlated with their star formation activity. Late-type galaxies are actively star-forming in general, while star formation in early-type galaxies tends to be much less active. Historically, the most common approach to classify galaxy morphologies was the {\it visual} inspection by human eyes (see \citealt{Kennicutt1998}). With the advent of  the large and digitalized format of astronomical datasets over the last decades (and accordingly the significant increase of the galaxy sample size), it has become much more common to use analytical (or automated) classification, such as S\'{e}rsic index ($n$, \citealt{Sersic1963}), concentration parameter ($C$, often defined as the ratio between half light radius and 90\% radius), stellar mass surface density, bulge-to-total ratio (e.g., \citealt{Strateva2001, Catinella2010,  Wuyts2011, Fang2013, Cook2019, Cook2020}).
A strong correlation between the morphologies and the \ion{H}{1} gas content of galaxies has also been well established---i.e., \ late-type galaxies are gas-rich, while early-type galaxies tend to be gas-poor. Indeed, in the past studies, the galaxy morphologies along the Hubble sequence were used to estimate their \ion{H}{1} gas mass (see \citealt{RH1994}), which is often used to quantify the ``deficiency'' of atomic hydrogen in galaxies residing in dense environments (\ion{H}{1} deficiency; \citealt{Gavazzi2008}, \citealt{Boselli2010}, \citealt{Hughes2013}). 

With the recent large, \ion{H}{1} 21 cm line survey in the local universe, such as the GALEX Arecibo SDSS Survey (GASS; \citealt{Catinella2013}) and the Arecibo Legacy Fast ALFA (ALFALFA) survey (\citealt{Haynes2018}), it has become possible to investigate the morphological dependence of the \ion{H}{1} gas fraction of local galaxies in a more statistical way. By using the data from the GASS survey and various morphological classifications from the visual classification to the automated methods, \citet{Calette2018} reported that late-type galaxies tend to have higher \ion{H}{1} gas fraction than early-type galaxies at a given stellar mass. While \citet{Calette2018} divided their sample into stellar mass bins, they did not split the sample by star formation rates (\textit{SFR}). Because late-type galaxies tend to have higher \textit{SFR}s than early-type galaxies in general and are expected to have a large gas reservoir, the morphological dependence reported in \cite{Calette2018} might be produced by the different \textit{SFR} between the early- and late-type galaxy samples. Moreover, some outliers do not follow the general trend between the average star formation activity and gas content or galaxy morphology, such as the \ion{H}{1}-excess systems (early-type galaxies with large \ion{H}{1} reservoirs and negligible star-formation; \citealt{Gereb2016, Gereb2018}) or passive spirals (e.g., \citealt{George2017, Guo2020} but see also \citealt{Cortese2012}).

A more recent study by \cite{Cook2019} used extended GALEX Arecibo SDSS Survey (xGASS; \citealt{Catinella2010, Catinella2018}) to investigate the relation between \ion{H}{1} gas mass and the bulge-to-total ratio (measured with the 2-D Bayesian light profile fitting code (ProFit); \citealt{Robotham2017}) for local galaxies. They showed that the \ion{H}{1} gas mass of star-forming galaxies do {\it not} depend on the bulge-to-total mass ratio, suggesting that the presence of bulge has little impact on their \ion{H}{1} gas content.
However, we should note that \cite{Cook2019} selected star-forming galaxies by simply excluding the quiescent galaxy population based on the distance from the star formation main sequence \citep{Catinella2018, Janowiecki2019}. Also, their morphological classification is different from that of \citet{Calette2018}. In these ways, the sample definition and/or the morphological classifications vary from study to study, making it difficult to interpret the impact of galaxy morphologies on their \ion{H}{1} gas content.

In our study, we carry out a comprehensive analysis by using the multiple morphological indicators (S\'{e}rsic index, C-index, and visual smoothness) and by dividing our sample by their stellar mass and \textit{SFR}. We stack the radio spectra of galaxies from the ALFALFA survey to calculate the average gas fraction. Then we can compare the average \ion{H}{1} gas content of galaxies with different morphologies \textit{at the same stellar mass and star formation rate.}

The structure of this paper is as follows. In Sec. \ref{sec:data}, we show our sample selection and summarize the galaxy information. The procedure to stack the radio spectra and the stacking result for our ``star-forming'' galaxy sample are summarized in this section. We then investigate the morphological dependence of \ion{H}{1} gas mass at fixed stellar mass and \textit{SFR} in Sec. \ref{sec:morph}, and interpret our results in Sec. \ref{sec:discussion}. Finally, we summarize our results in Sec. \ref{sec:summary}.
Throughout this paper, we adopt a flat $\Lambda$CDM cosmology with $\Omega_m=0.3$, $\Omega_\lambda=0.7$, and $H_0=67.8\ {\rm km\ s^{-1}\ Mpc^{-1}}$ and Chabrier initial mass function \citep{Chabrier2003}.

\section{Data and \ion{H}{1} spectral stacking} \label{sec:data}

To investigate the morphology dependence of \ion{H}{1} contents in local galaxies, we need not only \ion{H}{1} information but also various parameters of host galaxies. A combination of the large data sets from GALEX-WISE-SDSS Legacy Catalog 2 (GSWLC2, \citealt{Salim2018}) and ALFALFA survey enables us to study the \ion{H}{1} gas properties of galaxies statistically.

Our sample is drawn from the GSWLC2, which contains the stellar mass and the star formation rate of $\sim700,000$ galaxies derived from the fitting of Spectral Energy Distribution for the multiwavelength dataset. We then cross-matched the GSWLC2 catalog with the MPA-JHU catalog \citep{Abazajian2009} of SDSS DR7 to obtain the spectroscopic redshift of each target. We then select 32,717 galaxies (with $0.01<z<0.05$) located within the ALFALFA \ion{H}{1} 21 cm line survey footprint. Here we note that 14,252 galaxies having companions within the cut-out region of radio spectra (4 arcmin; see Sec.\ref{subsec:stacking}) are excluded to avoid overestimating \ion{H}{1} gas mass and the effect of galaxy-galaxy interaction.
To identify companions, we do not apply any criteria for stellar mass, color, or redshift. This method increases the number of discarded galaxies that are potentially not accompanied. However, in this study, we conservatively discard all galaxies having any companion within the cut-out region of \ion{H}{1} emission line.
In the following analysis, we restrict the sample to 18,465 galaxies with $M_*$ and \textit{SFR} range of $9.0<\log_{10}M_*<12.0$ and $-2.0<\log_{10}SFR<2.0$. In addition, considering the possible effect of active galactic nuclei (AGN) in our analysis, we exclude 7,573 AGN host candidates from our sample using the BPT diagram \citep{Kauffmann2003}. Here we used the [\ion{O}{3}], [\ion{N}{2}], H$\alpha$ and H$\beta$ emission line fluxes of individual galaxies from the MPA-JHU catalog, while we exclude 5 galaxies for which any of the emission line fluxes are not properly measured in the MPA-JHU catalog.
In this study, we use the remaining 10,887 galaxies as our parent `star-forming' galaxy sample.

\subsection{\ion{H}{1} spectral stacking} \label{subsec:stacking}

The \ion{H}{1} gas mass, $M_{{\rm H_I}}$, is derived from the 21 cm line flux \citep{Roberts1963} in this paper. Although the \ion{H}{1}-detected source catalog from the ALFALFA survey is already public (\citealt{Haynes2018}), directly using the \ion{H}{1}-detected source catalog may cause a potential bias towards the \ion{H}{1}-rich galaxies due to the detection limit.
The \ion{H}{1} gas mass, $M_{{\rm H_I}}$, is derived from the flux of 21 cm line ($S$) following the equation \ref{eq:HIFtoM} (Roberts 1963),
\begin{eqnarray}
\frac{M_{{\rm H_I}}}{M_\odot}=\frac{2.356\times10^5}{1+z}\left[\frac{D_L(z)^2}{{\rm Mpc}}\right]\left(\frac{S}{{\rm Jy\ km\ s^{-1}}}\right),\label{eq:HIFtoM}
\end{eqnarray}
where $D_L(z)$ is the luminosity distance to the object. The lower limit of detectable \ion{H}{1} gas mass depends on the distance to the object. More distant galaxies with less \ion{H}{1} gas, therefore, might not be detected through the pipeline of ALFALFA survey.

To avoid this potential bias, we apply the \ion{H}{1} stacking analysis following the previous studies (e.g., \citealt{Fabello2011}, \citealt{Brown2015}).
The radio spectra around \ion{H}{1} 21 cm line of our sample ($8'\times8'$ square) are extracted from the full volume ALFALFA data cubes using R.A., Dec, and $cz$ in the MPA-JHU catalog regardless of the detection of \ion{H}{1} 21 cm line. A spectrum of one galaxy has two polarizations and contains velocity, flux, and quality weight $w$ for each velocity bin. The quality weight $w$ evaluates the effect of radiofrequency interference and/or hardware issues, ranging from 0 (unusable data) to 1 (good data). We discard velocity bins with $w<0.5$ and take the average of two polarizations. If more than 40\% of velocity bins have $w$ less than 0.5 in a spectrum, we exclude such spectrum from our stacking analysis. Forty galaxies are removed through this procedure. After that, the spectrum is rest-frame shifted using the redshift from optical spectroscopy so that the center of \ion{H}{1} 21 cm line should be set to $0$ km/s. Also the noise level of each spectrum is evaluated using $rms$ of the spectrum outside of \ion{H}{1} emission from the target galaxy ($-1000<v\ [{\rm km\ s^{-1}}]<-500$ and $500<v\ [{\rm km\ s^{-1}}]<1000$, see Sec. \ref{subsec:ALLstack}). Because we eliminate the sources having one or more close companions, our sample can be free from the effect of the beam confusion \citep{Jones2016}.

By considering the possible correlations between \ion{H}{1} gas properties, stellar mass, and the \textit{SFR}, we aim to study the morphological dependence of \ion{H}{1} gas properties {\it at fixed stellar mass and SFRs}.
In the stacking process, we give a weight for each spectrum as $(1+z)^2/D_L(z)^2M_*$ (see Sec. 3.3 in \citealt{Fabello2011}). If we stack the radio spectra without any weight, massive and/or nearby galaxies will significantly contribute to the resulting \ion{H}{1} gas fraction ($F_{\rm H_I}$). In equation \ref{eq:HIFtoM}, $1+z$ is not squared because it assumes that \ion{H}{1} 21 cm line flux $S$ is rest-frame shifted correctly. However, because our calculation starts from the observed \ion{H}{1} 21 cm line flux $S_{obs}$, we have to weight $1+z$ in the rest-frame shifting. We solve this problem by changing the stacking weight from $(1+z)/D_LM_*$ to $(1+z)^2/D_LM_*$. We also note that we perform stacking analysis only for the bins with a sample size of $\geq$10.

\begin{figure}
\plotone{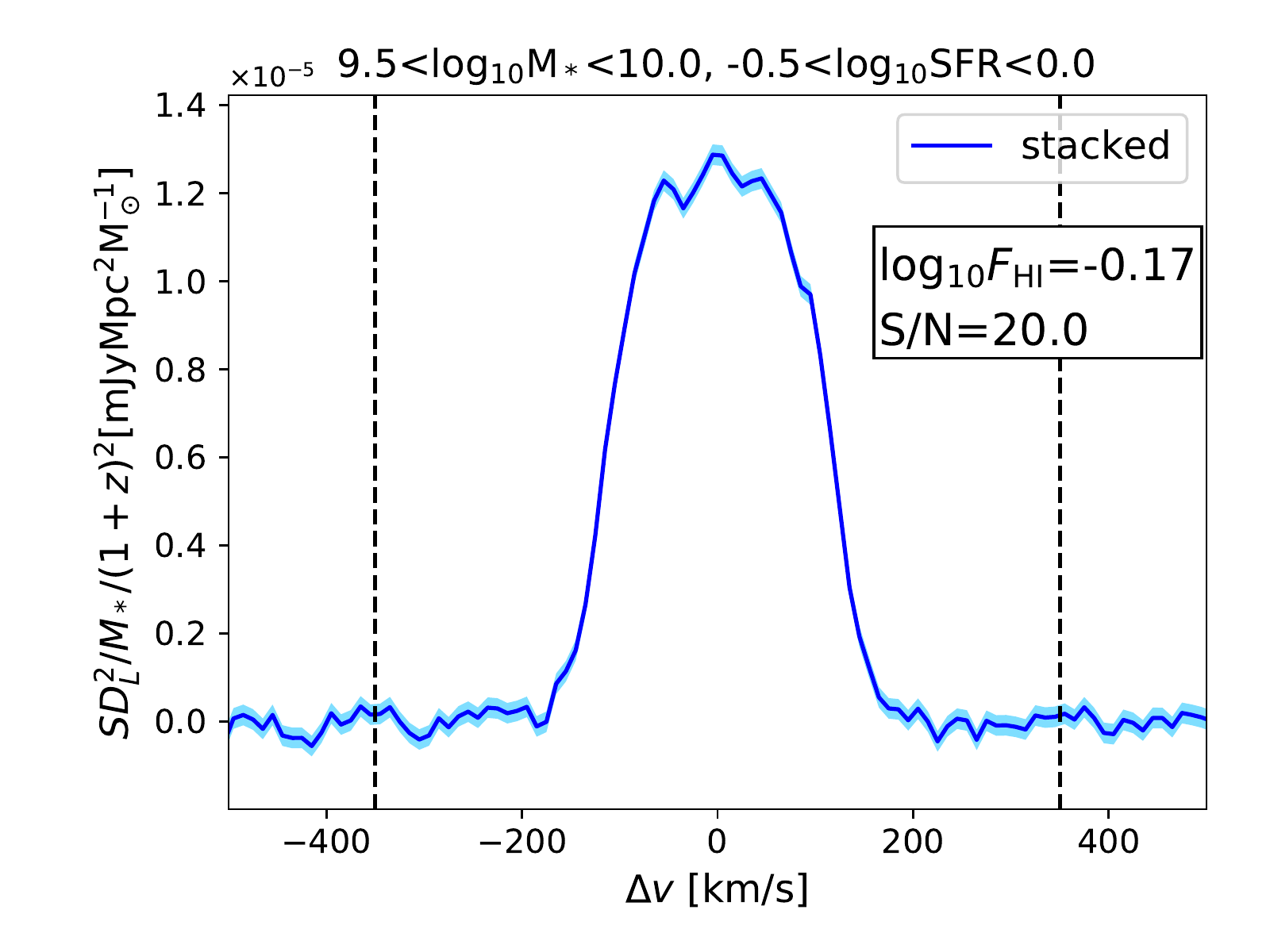}
\caption{An example of the stacked \ion{H}{1} spectra obtained in Sec. \ref{subsec:ALLstack}. The light-blue region shows the error in each velocity bin derived from the $rms$ of individual spectra. Vertical dashed lines are shown at $\pm350$ km/s, between which the flux is summed up to calculate the gas fraction ($F_{\rm H_I}$, see Sec. \ref{subsec:stacking}).
\label{fig:spec}}
\end{figure}

After subtracting the baseline of a stacked spectrum by linear fitting, we calculate the average gas fraction in each bin by summing up the flux in the velocity range of $-350<v\ [{\rm km\ s^{-1}}]<350$ (dashed lines in Figure \ref{fig:spec}). This velocity range is determined to cover the widest emission line after the stacking process to be performed in Sec. \ref{subsec:ALLstack}. Adopting a fixed velocity range may lead to an increase in the error of $F_{{\rm H_I}}$ measurements.
However, we decided to adopt this fixed velocity range for all the stacked spectra to estimate the errors and upper limits for \ion{H}{1} undetected sources in a consistent manner. The flux error of each velocity component in a stacked spectrum is derived by propagating the $rms$ from individual spectra before stacking (the light-blue region in Figure \ref{fig:spec}).
On the other hand, we adopt the Bootstrap resampling approach when calculating the average gas fractions to consider the sampling errors. In the following analyses, we use the average of gas fraction $F_{\rm H_I}$ and its standard deviation from 1000 times Bootstrap resampling as the average gas fraction and its $1\sigma$ error. Here we note that the sampling error dominates the total error budget in our analyses.

\subsection{Average \ion{H}{1} gas fraction across the star-forming main sequence}\label{subsec:ALLstack}

\begin{figure}
\plotone{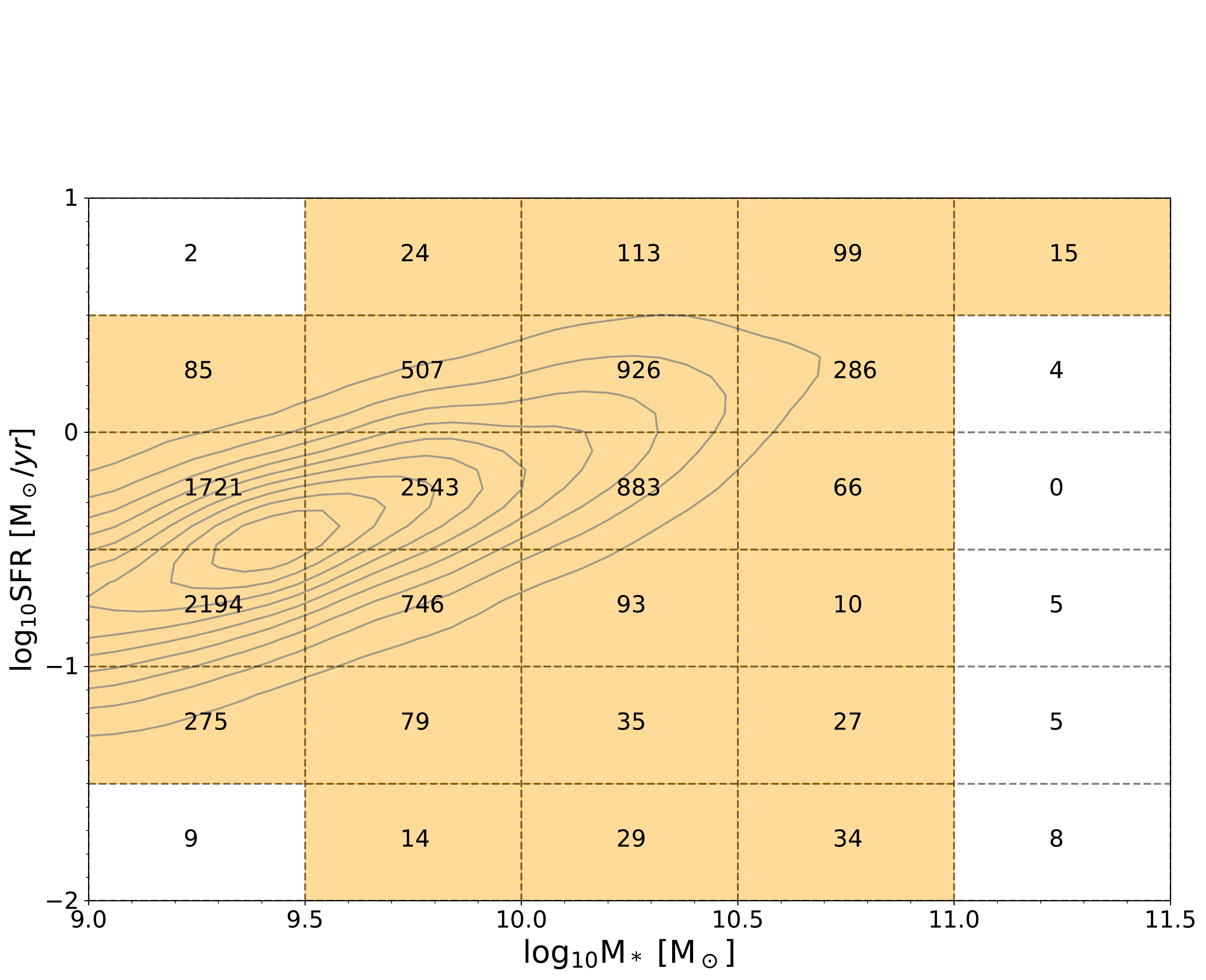}
\caption{The $M_*$-\textit{SFR} diagram to show the distribution of all our sample (grey contours) and the number of galaxies used for \ion{H}{1} stacking at each ($M_*$, \textit{SFR}) bin. The grid size indicates  $\Delta M_* = \Delta SFR = 0.5$ dex (see Sec. \ref{subsec:ALLstack} for details). Orange shaded bins are used for the stacking (N$\geq$10).
\label{fig:Ndis_ALL}}
\end{figure}

Firstly, we simply divide our sample by the stellar mass and \textit{SFR} ($\Delta M_* = \Delta \mathit{SFR} = 0.5$ dex), and carry out the stacking process as explained above. 
Figure \ref{fig:Ndis_ALL} summarizes the number of galaxies included in each bin.
The sample size tends to be small at high $M_*$ and/or low \textit{SFR} ends, and we do not perform \ion{H}{1} stacking analyses for the bins with the sample size of $N<10$. We note that the results presented in this paper mainly focus on galaxies on the star-forming main sequence.

\begin{figure*}
\plottwo{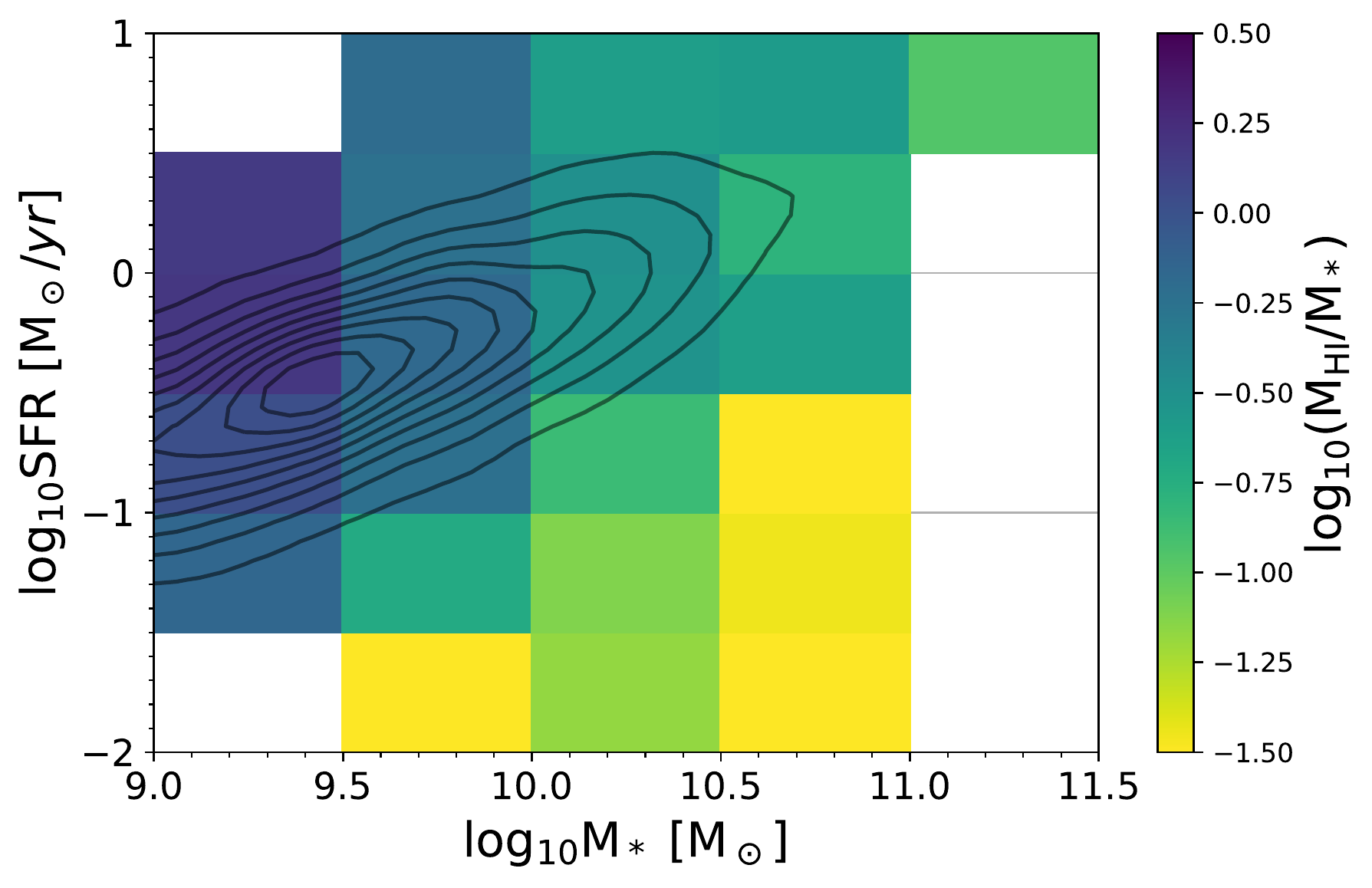}{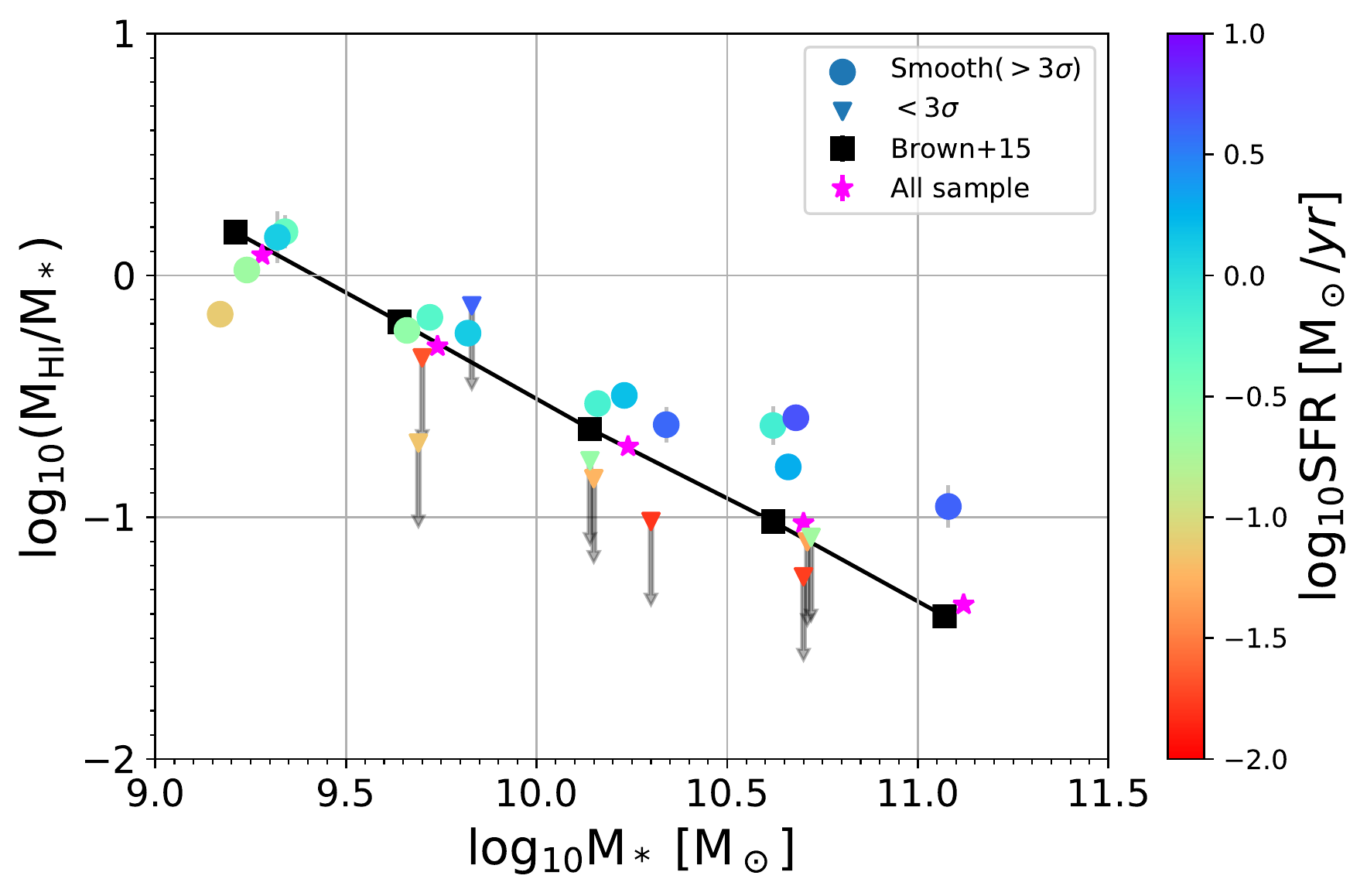}
\caption{{\bf Left:} The average gas fraction (see color bar) at each position on the stellar mass vs. SFR plane from our ALFALFA stacking analysis.
We include the marginal detection ($S/N<3$) in this panel. The gas fraction depends both on the stellar mass and \textit{SFR}. The black contours show the distribution of the parent `star-forming' galaxies.
{\bf Right:} The gas fraction plotted against stellar mass. The color-coding of this plot shows the median \textit{SFR} of each bin. Circles are the points with $S/N>3$, and inverted triangles mean the $3\sigma$ upper limit.
The magenta stars show the stacking result for all local galaxies by simply dividing our sample into five stellar mass bins (without considering \textit{SFR} difference). Here, we include the galaxies having the potential companion within four arcmins to set the same selection criteria to \citet{Brown2015}. The gray error bars are derived from the Bootstrap method (Sec. \ref{subsec:stacking}). We show the $M_*$--$F_{{\rm H_I}}$ scaling relation derived by \citet{Brown2015} as the black squares connected by the black solid line.} \label{fig:ALLstack}
\end{figure*}

In the left panel of Figure \ref{fig:ALLstack}, the color-coding indicates the average gas fraction of galaxies at each stellar mass and \textit{SFR}.
Here we include the marginal detection down to $S/N=3$.
It can be seen that the average gas fraction gradually changes with the stellar mass and the \textit{SFR}. The increasing gas fraction with increasing \textit{SFR} at fixed stellar mass seen in this diagram is consistent with a  picture that the rich gas reservoir supports an active star-formation \citep{Tumlinson2017}.

In the right panel of Figure \ref{fig:ALLstack}, we show the relation between the gas fraction and stellar mass, with the color-coding based on the median \textit{SFR} in each bin.
The upper limits represented by the triangles and arrows are the $3\sigma$ value.
It is clear from the right panel of Figure \ref{fig:ALLstack} that there is a strong correlation between \ion{H}{1} gas fraction and stellar mass, consistent with previous studies.
The black squares connected by the solid line in Figure~\ref{fig:ALLstack} (right) show the \ion{H}{1} gas scaling relation derived by \citet{Brown2015}. Our results are in good agreement with the black lines at $M_*<10^{10.2}M_\odot$, while at the massive end, our results tend to show higher values than the average scaling relation determined by \citet{Brown2015}.
The distinction of $F_{gas}$ in the massive side is not surprising because we performed the \ion{H}{1}  stacking analyses only for the bins with the sample size of $N\geq10$, and most of the bins below the star-formation main sequence are not used in this study.
\citet{Brown2015} suggested that star-forming galaxies tend to host more \ion{H}{1} gas than quenched galaxies, which can explain the difference between \citet{Brown2015} and our results at $M_*>10^{10.2}M_\odot$. We emphasize that our stacking analysis shows an excellent agreement with \citet{Brown2015} when we perform the \ion{H}{1} stacking using all SDSS galaxies at each stellar mass bin without considering the \textit{SFR} binning (see magenta stars in Figure \ref{fig:ALLstack} right).

\section{Morphological dependence of \ion{H}{1} gas fraction at fixed $M_*$ and \textit{SFR}}\label{sec:morph}

The primary goal of this study is to investigate the {\it morphological} dependence of the \ion{H}{1} gas fraction of our sample \textit{at fixed stellar mass and SFR}. Here we use three morphological indicators of galaxies; S\'{e}rsic $n$, concentration index (C-index), and visual smoothness from the Galaxy Zoo 2 project (GZ2, \citealt{Lintott2008, Willett2013, Hart2016}).

The S\'{e}rsic profile can reproduce the classical surface brightness distribution of early-type galaxies (the de Vaucouleurs profile) when $n\sim4$, and the exponential disk profile of late-type galaxies when $n=1$. 
The C-index is often defined as the ratio of the half-light radii $R_{50}$ and $R_{90}$, where $R_x$ means the radius in which $x$ \% of the total flux is enclosed. It is shown that the C-index has a strong correlation with the dominance of the bulge component in galaxies (e.g., \citealt{Shimasaku2001}). We also introduce the ``visual smoothness'' from GZ2 as a proxy for the traditional visual classifications made by human eyes (see Sec.~\ref{subsec:GZ2} for more details).

We cross-matched our sample constructed in Sec.\ref{subsec:ALLstack} with the New York University Value-Added Catalog (NYU-VAGC; \citealt{Blanton2005}) and the full GZ2 catalog \citep{Hart2016}.
This new catalog contains 7712 objects, all of which have measurements of $M_*$, \textit{SFR}, redshift, ALFALFA radio spectra, S\'{e}rsic $n$, C-index, and visual smoothness.
We discuss the morphological impacts on the $M_*$--$F_{{\rm H_I}}$ relation with this new sample in the following subsections.

\subsection{S\'{e}rsic index}\label{subsec:sersic}

\begin{figure}
\plotone{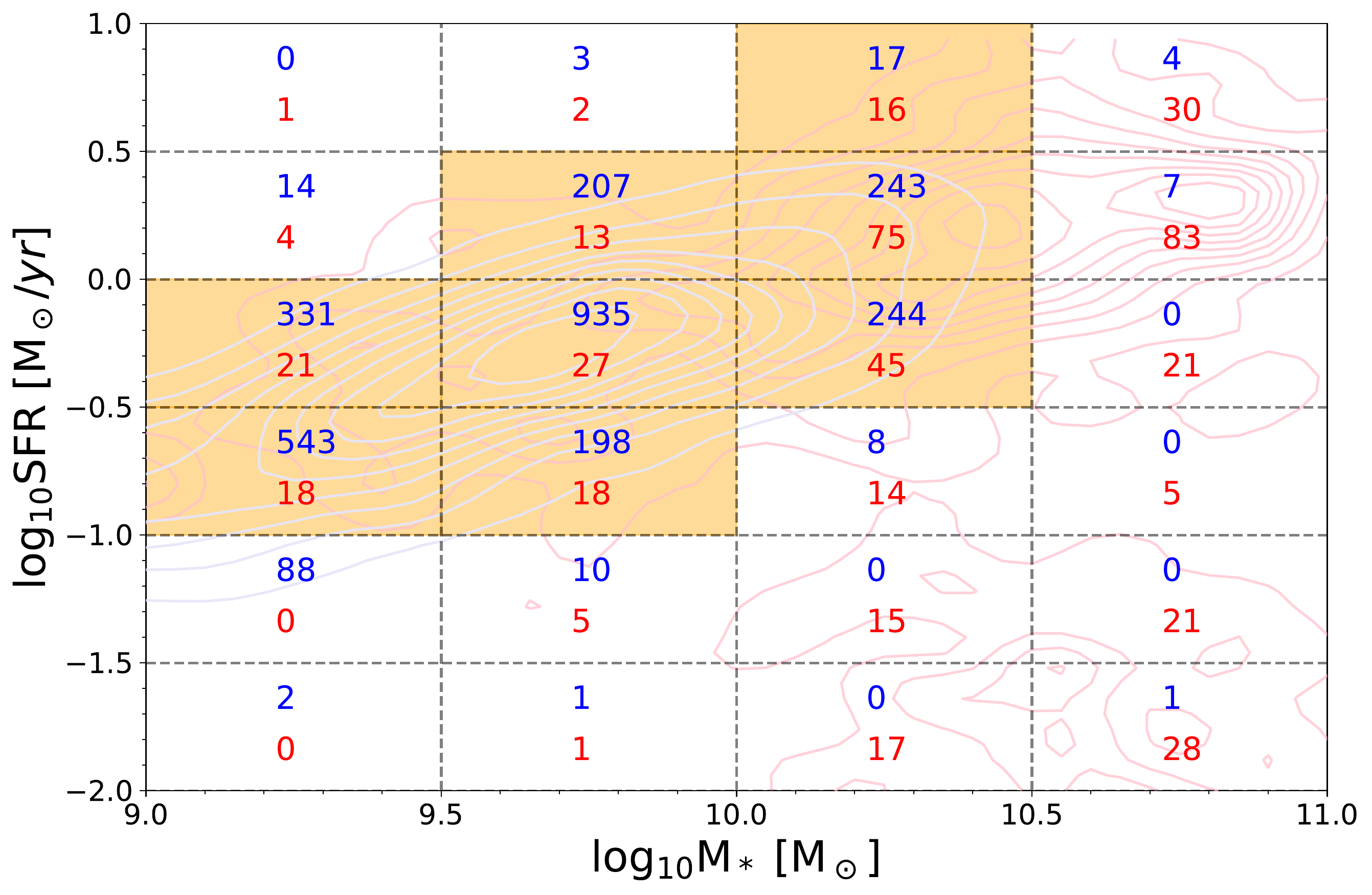}
\caption{Same figure as Figure \ref{fig:Ndis_ALL} but for late-type (blue) and early-type (red) galaxies divided by the S\'{e}rsic $n$. Blue and red contours show the distribution of late-type and early-type galaxies on this plane, respectively. Here we normalize the contours based on the total number of each population. The width of the grid is $\Delta M_* = \Delta SFR = 0.5$ dex. Here we show the region of $9.0<\log_{10}M_*<11.0$ and $-2.0<\log_{10}SFR<1.0$ because all other bins have sample sizes of $N<10$. Orange shaded bins are used for the comparison of $F_{\rm HI}$ in panel (c) of Figure~\ref{fig:comp_morph} ($N\geq10$ for both populations).
\label{fig:Ndis_Sersic}}
\end{figure}

We use the S\'{e}rsic index $n$ measured in the $r$-band of individual galaxies published in NYU-VAGC. In this study, we select 498 early-type and 2856 late-type galaxies by applying $n>3.5$ and $n<1.5$, respectively. Red and blue contours in Figure~\ref{fig:Ndis_Sersic} show the distribution of early-type and late-type galaxies on the $M_*$-\textit{SFR} diagram. Here we normalize the contours based on the total number of each population.
Because we mainly focus on the galaxies on the star-formation main sequence, the number of late-type galaxies is larger than that of early-type galaxies.

We then conducted the stacking analysis to the early/late-type subsamples in the same way as described in Sec.\ref{subsec:ALLstack}. Here we use the same bin size for stacking as that in Sec.\ref{subsec:ALLstack} ($\Delta M_*=\Delta \mathit{SFR}=0.5$ dex).
The number of early/late-type galaxies in each bin is shown in Figure \ref{fig:Ndis_Sersic}.

\begin{figure*}
\plotone{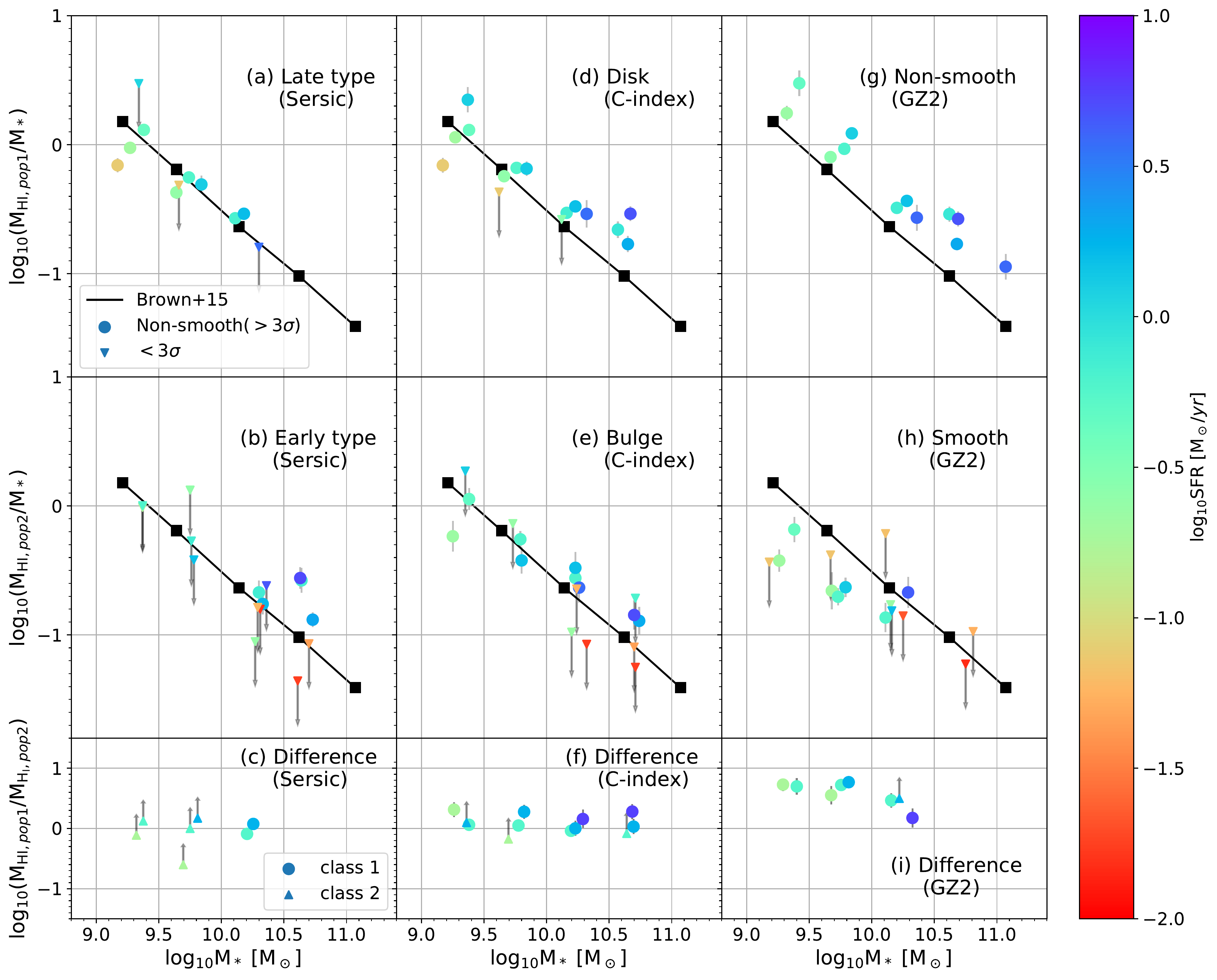}
\caption{The result of our stacking analysis when we divide our sample by S\'{e}rsic $n$ (left), C-index (middle), and visual smoothness (right). The top panels (a)/(d)/(g) show the relation between the stellar mass and the average \ion{H}{1} gas fraction of late-type/disk/non-smooth populations, respectively. The second row (b)/(e)/(h) is the same as the upper panels but for early-type/bulge/smooth galaxies.
The meanings of the large circles and inverted triangles in the top and second row are the same as the right panel of Figure \ref{fig:ALLstack}.
The panels (c)/(f)/(i) show the difference in the amount of \ion{H}{1} gas between the above two populations as a function of their stellar mass. 
The meaning of each symbol is summarized in Sec. \ref{subsec:sersic}.
Colors indicate the median \textit{SFR} of each subsample. The gray error bars are also derived by the Bootstrap method (see Sec.\ref{subsec:stacking}).
The black squares connected by the solid line show the scaling relation from \citet{Brown2015} (same as Figure \ref{fig:ALLstack}). \label{fig:comp_morph}}
\end{figure*}

The left column in Figure \ref{fig:comp_morph} show the result. Panel (a) and (b) in Figure \ref{fig:comp_morph} show the relationship between the median stellar mass and the average \ion{H}{1} gas fraction when we divide our sample into late-type and early-type by the S\'{e}rsic $n$. As we did in Figure \ref{fig:ALLstack}, we use the bins with $N\geq10$ for each morphological type. The general trend does not change from Figure \ref{fig:ALLstack}; the \ion{H}{1} gas fraction is anticorrelated with the stellar mass and correlated with the \textit{SFR}.

We compute the difference of \ion{H}{1} gas mass between early-type and late-type galaxies with the same stellar mass and \textit{SFR}, and they are plotted in Figure \ref{fig:comp_morph} (c) as a function of stellar mass. This comparison is performed for the bins in which both populations have more than nine galaxies in Figure \ref{fig:Ndis_Sersic}. 
Here we note that there are two classes of bin types in terms of the \ion{H}{1} detection.
\begin{enumerate} 
\item Both populations have $S/N>3$ ($\circ$ in Figure \ref{fig:comp_morph}).
\item The $F_{{\rm H_I}}$ of late-type/disk/non-smooth galaxies show $S/N>3$, but early-type/bulge/smooth galaxies show $S/N<3$ ($\triangle$ in Figure \ref{fig:comp_morph}).
\end{enumerate}
There are no bins with properties opposite to those of class 2 for all morphological indicators. In Figure \ref{fig:comp_morph} (c), we use large circles and normal triangles for the class 1 and 2, respectively. The number of bins in each class is summarized in Table \ref{tab:Nbin}.

Unfortunately, there are only two bins in class 1, and the remaining bins are class 2 or non-detection ($S/N<3$). The $F_{gas}$ differences in these two class 1 bins are $-0.09$ and $0.07$, suggesting that there is no significant difference in the \ion{H}{1} gas mass between late-type and early-type galaxies at fixed $M_*$ and \textit{SFR} in those bins.
Outside of these two bins ($\log_{10}M_*<10.0$ or $\log_{10}M_*>10.5$), we could not detect \ion{H}{1} in the stacked spectra due to the poor statistics, especially for early-type galaxies.

We also conduct the same analysis with the different S\'{e}rsic $n$ criteria. Here early/late-type galaxies are defined as those sitting in the top/bottom 20\%-ile profile in the distribution of S\'{e}rsic $n$; i.e.\ $n<1.25$ for late-types and $n>2.30$ for early-types. This method enables us to match the total size of subsamples and increase the number of bins in class 1. In the range of $9.5<\log_{10}M_*<10.5$ and $-0.5<\log_{10}SFR<0.5$, we find that there is no significant difference in the \ion{H}{1} gas mass between early- and late-type galaxies. Using this method, however, early-type subsample can include contaminants from ``intermediate'' morphology ($n\sim2-3$). It is, therefore, difficult to conclude there is no difference in HI gas mass fraction between early-type and late-type galaxies at fixed stellar mass and \textit{SFR} with this percentile approach.

\begin{deluxetable}{ccccccc}
\tablenum{1}
\tablecaption{The number of bins for class 1 and 2 according to the S/N of the stacked \ion{H}{1} flux}.\label{tab:Nbin}
\tablewidth{0pt}
\tablehead{
\colhead{} & \colhead{} & \colhead{} & \colhead{class} 1 & \colhead{} & \colhead{} & \colhead{class 2} \\\hline
\colhead{$S/N_{\rm pop.1}(F_{{\rm H_I}})$} & \colhead{} & \colhead{} & \colhead{$>3$} & \colhead{} & \colhead{} & \colhead{$>3$}\\
\colhead{$S/N_{\rm pop.2}(F_{{\rm H_I}})$} & \colhead{} & \colhead{} & \colhead{$>3$} & \colhead{} & \colhead{} & \colhead{$<3$}
}
\startdata
S\'{e}rsic $n$ & & & 2 & & & 5\\
C-index & & & 9 & & & 3\\
Visual smoothness & & & 7 & & & 1\\
\enddata
\tablecomments{The ``class 1'' means that \ion{H}{1} is detected at $S/N>3$ for both of the populations in that bin. The ``class 2'' indicates the case where the first population shows $S/N(F_{{\rm H_I}})>3$, while the second population has $S/N(F_{{\rm H_I}})<3$. There are no bins with the properties opposite to those of class 2. In the lower row, we show the number of bins for each class and each morphological index. The class 1 and 2 are shown as circles and triangles in the bottom panels of Figure \ref{fig:comp_morph}, respectively.}
\end{deluxetable}

\subsection{Concentration index}\label{subsec:cindex}
\begin{figure}
\plotone{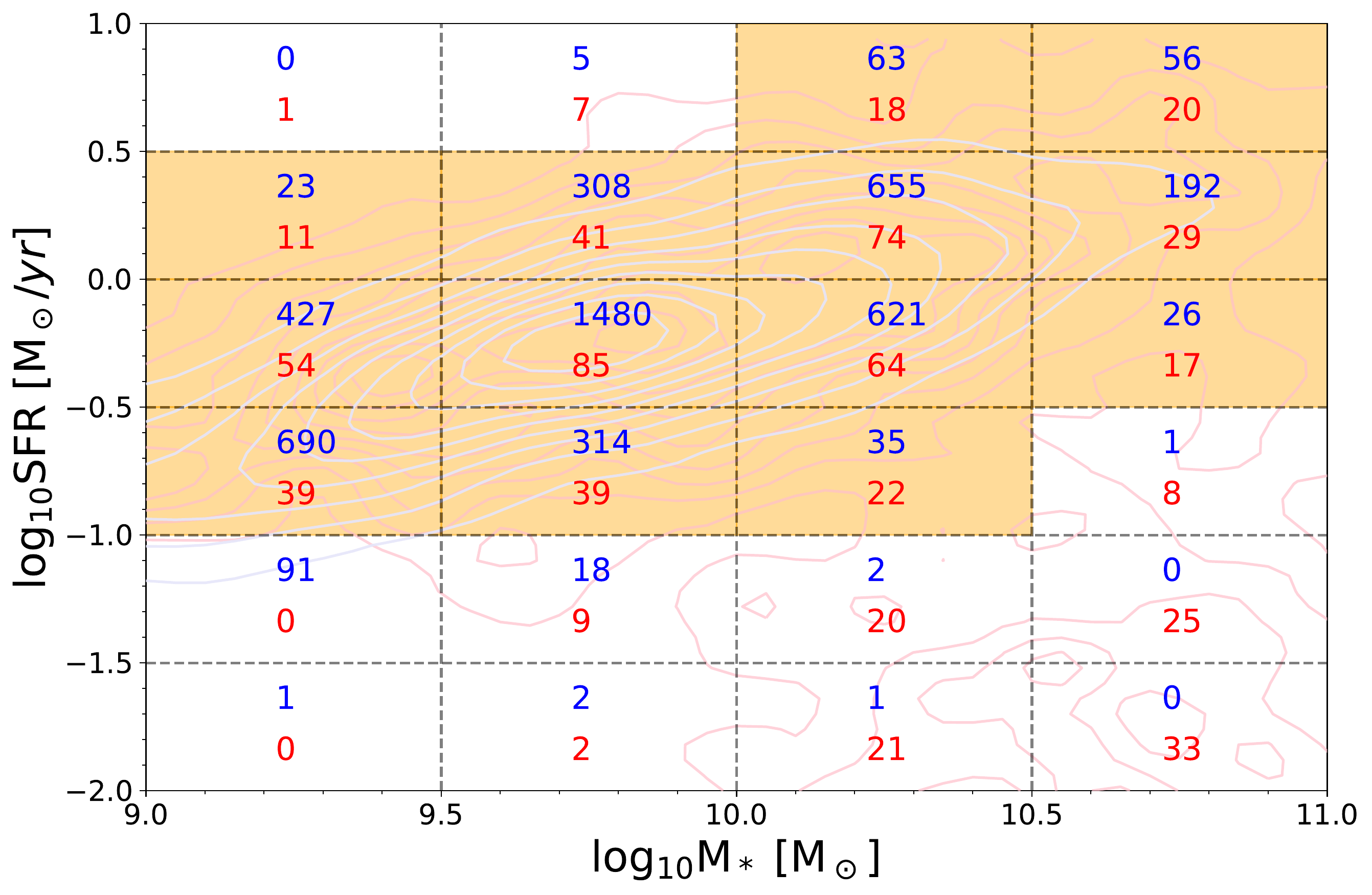}
\caption{Same figure as Figure \ref{fig:Ndis_Sersic} but for disk (blue) and bulge (red) galaxies divided by the C-index.
\label{fig:Ndis_C}}
\end{figure}

We also retrieve the \textit{r}-band Petrosian half-light radii $R_{50}$ and $R_{90}$ from the NYU-VAGC catalog, and calculate C-index as a ratio of these two values ($R_{90}/R_{50}$). While the criteria of C-index used for the morphological classification are different from study to study (e.g., \citealt{Deng2013, Koyama2019}), we adopt $C>2.8$ and $C<2.5$ for bulge-dominated and disk-dominated galaxies, respectively. We obtain 660 bulge and 5020 disk galaxies in total. Figure~\ref{fig:Ndis_C} shows the $M_*$-\textit{SFR} diagram for bulge/disk galaxies. Binning width is the same as that in Sec.~\ref{subsec:sersic} ($\Delta M_*=\Delta \mathit{SFR}=0.5$ dex), and the number of galaxies in each bin is also summarized in Figure~\ref{fig:Ndis_C}.

The middle column of Figure \ref{fig:comp_morph} shows the results of our \ion{H}{1} stacking analysis when we use C-index for morphological classification. The panels (d) and (e) in Figure \ref{fig:comp_morph} show the scaling relation for disk- and bulge-dominated galaxies. In the panel (f), we show the difference between the disk and bulge samples by computing $M_{{\rm H_I,Disk}}$/$M_{{\rm H_I,Bulge}}$ at each stellar mass and \textit{SFR}. 
The meanings of the symbols are the same as those in panel (c). The weighted average of $\log_{10}(M_{{\rm H_I,Disk}}/M_{{\rm H_I,Bulge}})$ of the ``class 1'' bins ($S/N(F_{{\rm H_I}})>3$ for both bulge and disk populations) is $0.09\pm0.10$, suggesting that disk galaxies and bulge galaxies at fixed stellar mass and SFR have similar amount of \ion{H}{1} gas on the star-formation main sequence ($9.0<\log_{10}M_*<11.0$). We verify that our results are unchanged even if we instead use the upper/lower 20 percentile in the distribution of C-index to select bulge and disk populations ($C>2.59$ and $C<2.07$). Again, we note that the intermediate population can contaminate the morphological selection with this percentile approach.

\subsection{Visual classification with GZ2}\label{subsec:GZ2}

Galaxy Zoo project \citep{Lintott2008} provides {\it visual} classifications of morphologies given by {\it citizen} scientists for a large number of galaxies observed in SDSS. In the Galaxy Zoo project, citizen scientists are asked to inspect galaxies' images and choose their morphological class (e.g., elliptical, clock-wise spiral). Following Galaxy Zoo, Galaxy Zoo 2 project (GZ2, \citealt{Willett2013}) aims to describe more detailed morphological properties using a similar method to that of Galaxy Zoo.
We here focus on the first question of GZ2: {\it ``Is the galaxy simply smooth and rounded, with no sign of a disk?"}.
All participants of the GZ2 project are asked to choose one answer from ``Smooth and rounded'', ``Features or disk'', or ``Star or artifact''. Then, GZ2 calculates the ``vote fraction'' for each answer. \citet{Willett2013} mentioned that, for most of \textit{clean} spirals in Galaxy Zoo, the vote fraction for ``Features or disk'' in GZ2 is similar to that for spiral galaxies in Galaxy Zoo.

\citet{Hart2016} revised the vote fraction with the idea that the actual distribution of the vote fraction should be independent of the redshift in the local universe. They remove the effect of size, luminosity, and redshift of galaxies and introduce the {\it debiased} vote fraction. Following the recommendation by the GZ2 project, we use this {\it debiased} vote fraction from \citet{Hart2016} in this paper, rather than directly using the original vote fraction provided by \citet{Willett2013}.

\begin{figure}
\plotone{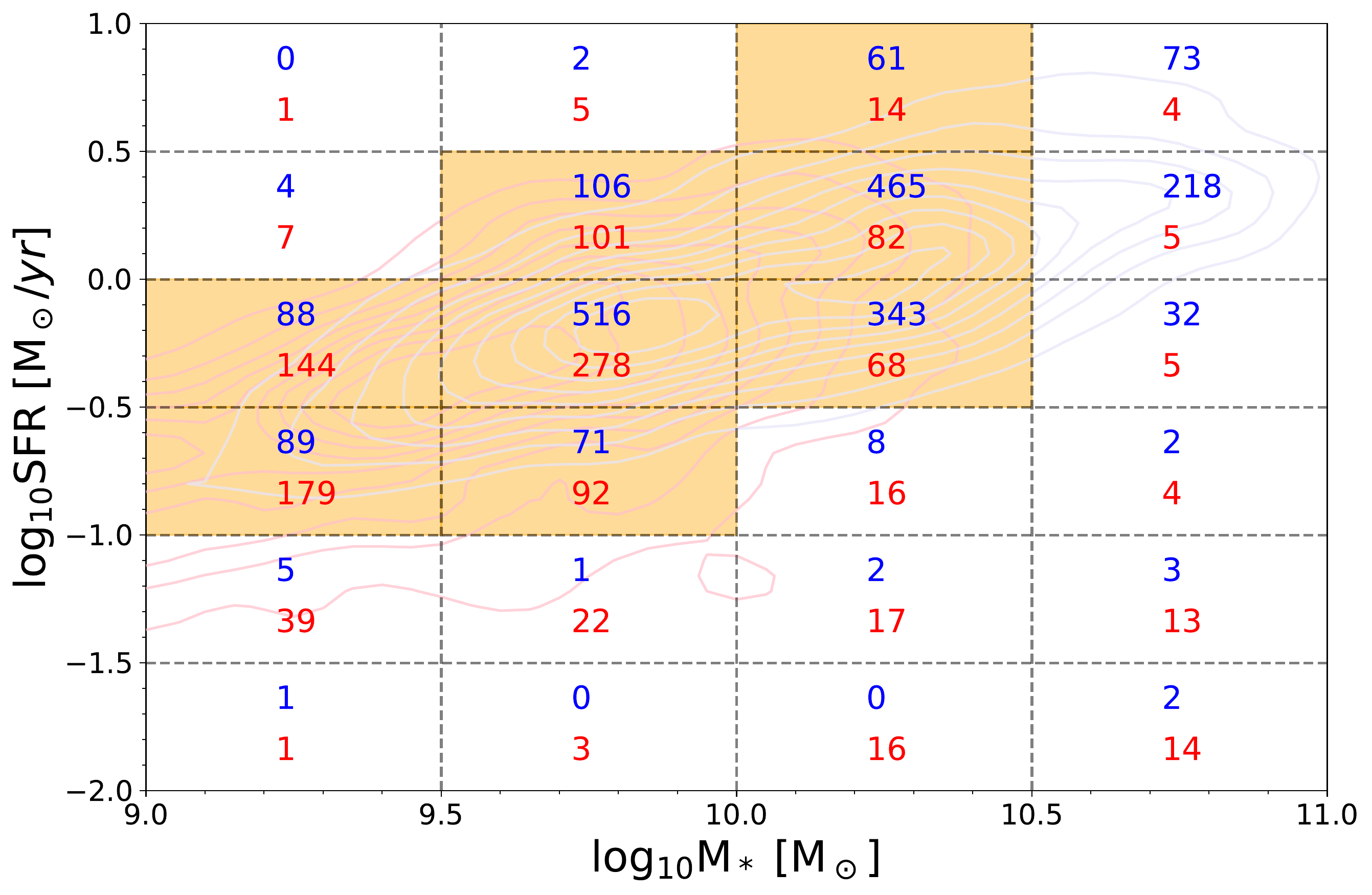}
\caption{Same figure as Figure \ref{fig:Ndis_Sersic} but for non-smooth (blue) and smooth (red) galaxies divided by the visual smoothness.
\label{fig:Ndis_GZ2}}
\end{figure}

We define ``smooth'' galaxies as those having the debiased vote fraction of $>0.8$ for ``Smooth and rounded'' in GZ2. On the other hand, we define ``non-smooth'' galaxies as those with the debiased vote fraction of $>0.8$ for ``Features or disk''. Here we follow the recommendation by \citet{Willett2013} to use the criteria of the debiased vote fraction $>0.8$ when performing any morphological classification with GZ2. The bin size for our \ion{H}{1} stacking analyses is the same as that in Sec \ref{subsec:sersic} ($\Delta M_*=\Delta \mathit{SFR}=0.5$ dex), and Figure \ref{fig:Ndis_GZ2} shows the number of galaxies in each bin.

The result is shown in the right column of Figure \ref{fig:comp_morph}. As we performed for S\'{e}rsic index and C-index in the previous subsections, we show in panels (g) and (h) the scaling relation for non-smooth and smooth galaxies, respectively. Panel (i) shows the difference between these two populations at fixed $M_*$ and \textit{SFR}. Interestingly, smooth galaxies tend to have significantly lower gas fraction than non-smooth galaxies, and we find that there are six bins where both non-smooth and smooth galaxies are detected at $S/N(F_{{\rm H_I}})>3$ (class 1). The weighted average of $\log_{10}(M_{{\rm H_I,Non-smooth}}/M_{{\rm H_I,Smooth}})$ is $0.71\pm0.11$, suggesting a significant difference in \ion{H}{1} gas fraction between the smooth and non-smooth galaxies at fixed stellar mass and \textit{SFR} in the range of $9.0<\log_{10}M_*<10.5$. This is the main result of our paper.
We find no bin where \ion{H}{1} is detected only for smooth galaxies. This result is also unchanged even if we use the percentile approach to define smooth and non-smooth galaxies as we did in Sec.\ref{subsec:sersic} and Sec.\ref{subsec:cindex}.

\section{Discussion} \label{sec:discussion}

The morphological dependence of the \ion{H}{1} gas mass in local galaxies has been discussed for decades (e.g., \citealt{RH1994}). The \ion{H}{1} gas mass is known to depend on the stellar mass and the \textit{SFR} of host galaxies (e.g., \citealt{Brown2015}). On the other hand, the morphology of galaxies is also related to the stellar mass and the \textit{SFR} (e.g., \citealt{Kauffmann2003}). These relationships between multiple parameters make it difficult to understand the real impact of galaxy morphologies on their \ion{H}{1} gas mass content.

In our study, we divide our sample into small bins on the $M_*$-\textit{SFR} plane ($\Delta M_*=\Delta SFR=0.5$ dex). The galaxies in each bin are further divided by their morphology with three morphological indicators; S\'{e}rsic $n$, C-index, and the visual smoothness.
Our study revealed that \textit{visually} smooth galaxies have lower gas fraction than non-smooth galaxies at fixed stellar mass and SFR.
Such a morphological trend is not observed when we use C-index (Figure \ref{fig:comp_morph}).
Below, we will discuss the potential candidates which may affect the \ion{H}{1} gas content of galaxies and why we see morphological dependence only when we use \textit{visual smoothness}.

\subsection{Stellar mass, \textit{SFR}, and environment} \label{subsec:MSFRdep}

When investigating the effect of galaxy morphology on the \ion{H}{1} gas mass, it is crucial to exclude the impact of other properties of galaxies. Many previous papers show that \ion{H}{1} gas mass strongly depends on the stellar mass of host galaxies ($M_*$-$M_{HI}$ scaling relation, e.g., \citealt{Fabello2011, Brown2015, Healy2019}). A simple approach to remove the stellar mass dependence is to check the \ion{H}{1} scaling relations of each subsample.
\citet{Calette2018} compared the \ion{H}{1} scaling relation of late-type galaxies and early-type galaxies and concluded that late-type galaxies have higher \ion{H}{1} gas fraction than early-type galaxies at all stellar mass range. However, the stellar mass is not the only parameter that determines the \ion{H}{1} gas mass of the host galaxies, and other parameters can also affect the scaling relation \citep{Cortese2011, Brown2015, Cook2019}. For example, \citet{Cook2019} studied the impact of the bulge-to-total mass ratio on the \ion{H}{1} scaling relation and found no significant effect of the morphology when they limited their sample to the star-forming galaxies. They argue that the morphological dependence of \ion{H}{1} scaling relation reported in previous studies originates from the correlation between the star-formation activity and galaxy morphology at fixed stellar mass. 
On the other hand, the amount of molecular hydrogen in the star-forming galaxy is shown to be approximately fixed in a wide range of bulge-to-total mass ratio, whereas that of \ion{H}{1} can change by a factor of 100 (see Figure 10 in \citealt{Catinella2018}).
This is also supported by \citet{Koyama2019}, who showed similar molecular gas mass fractions for green-valley galaxies with different morphologies (classified by C-index) at fixed stellar mass and \textit{SFR}.

Because all the analyses presented in this work are performed at fixed stellar mass and \textit{SFR}, our results should not be affected by different stellar masses or \textit{SFR}s. It is true that there remains a possibility that there is a small difference in the distribution of stellar mass and/or \textit{SFR} \textit{within} the small ($M_*$, SFR) bin.
We calculate the median $M_*$ and \textit{SFR} for smooth and non-smooth galaxies in each bin ($\delta M_*$ and $\delta SFR$), and confirm that the difference is very small for all the bins used in Figure \ref{fig:comp_morph} (i) ($\delta M_*\leq0.08$, and $\delta SFR\leq 0.11$). By recalling Figure \ref{fig:ALLstack}, the small difference existing in each bin cannot explain the large difference (0.71~dex) in the gas fraction between smooth and non-smooth galaxies.

Although we consider that the stellar mass and the star-formation activity are the primary parameters that determine the \ion{H}{1} gas content of galaxies, their surrounding environments may also affect the \ion{H}{1} scaling relation. Observations revealed that late-type galaxies in dense environments tend to have less \ion{H}{1} gas than those in the general field (e.g., \citealt{HG1984, Solanes2001}), likely due to environmental processes such as galaxy-galaxy interactions or ram pressure stripping \citep{Moore1998, GG1972, Cortese2021}. It is also true that early-type galaxies in the traditional Hubble sequence (ellipticals/S0s) tend to be located in cluster environments in the local universe (e.g., \citealt{Dressler1980}, \citealt{Bamford2009}).

To check the environment of smooth and non-smooth galaxies, we investigate the halo mass of smooth and non-smooth galaxies. By cross-matching our sample with the SDSS DR7 group catalog from \citet{Yang2008}, we obtain the halo mass of galaxies. We then divide these galaxies into small bins of stellar mass and \textit{SFR} as used for our stacking analysis ($\Delta M_*=\Delta \mathit{SFR}=0.5$ dex), and carry out the Kolmogorov–Smirnov (KS) test on the halo mass distribution of two populations. We find that $p$-values are $>$0.05 for most of the bins, suggesting that we cannot rule out the null hypothesis that the two populations are drawn from the same parent population. Therefore, it is unlikely that our results shown in Figure \ref{fig:comp_morph} are affected by the dense environment.

We note again that the morphological difference in the \ion{H}{1} gas fraction is visible only when we use \textit{visual smoothness} for the morphological classification. We do not see the difference when we use C-index (Figure \ref{fig:comp_morph} (f)), consistent with the conclusion of \citet{Cook2019}. A question here is---what is the {\it visual smoothness}? We will discuss this in Sec. \ref{subsec:vsmooth}, and attempt to identify reasons responsible for the different behavior when we use {\it smoothness} as a morphological indicator.

\subsection{Visual smoothness}\label{subsec:vsmooth}

\begin{figure}
\plotone{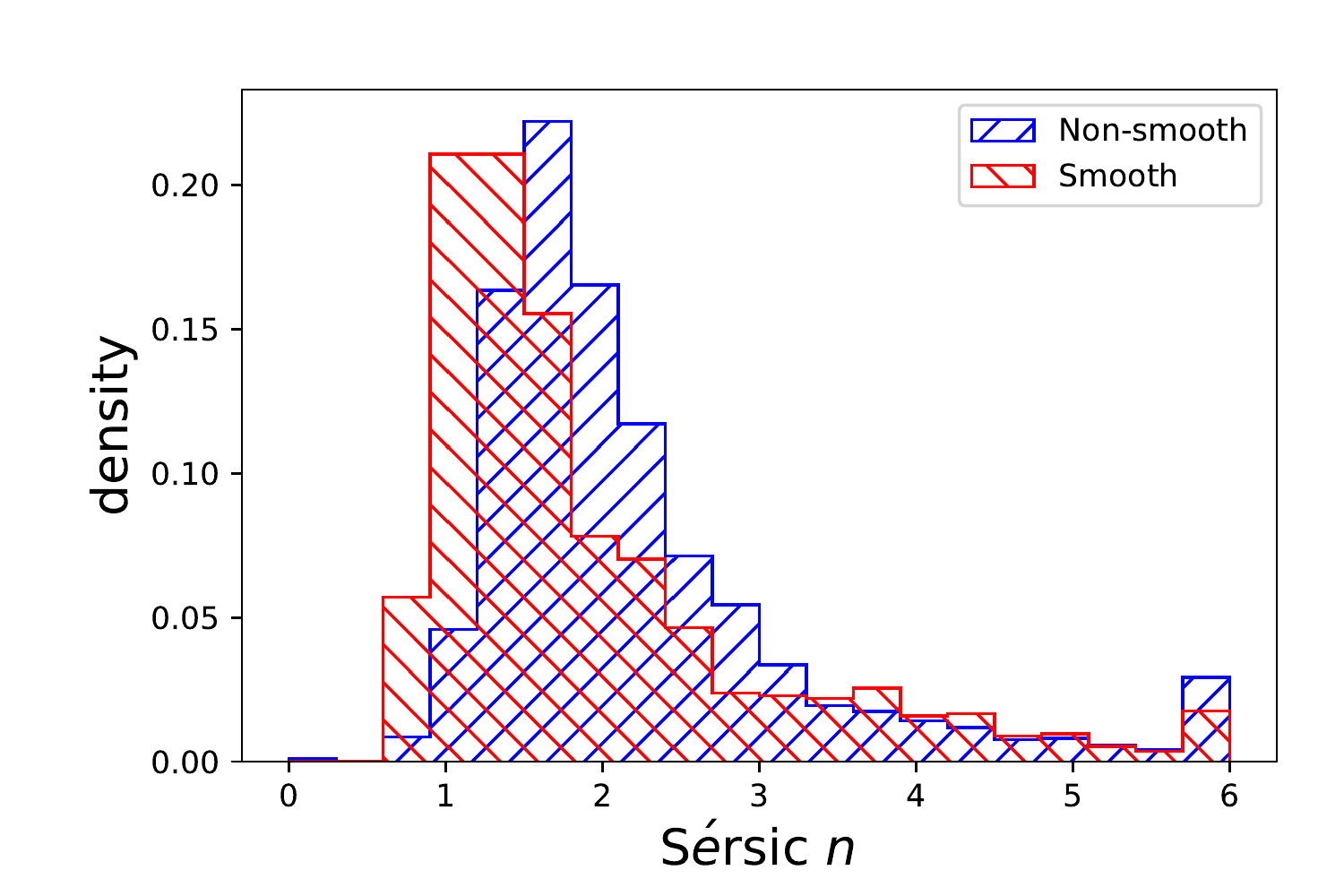}
\plotone{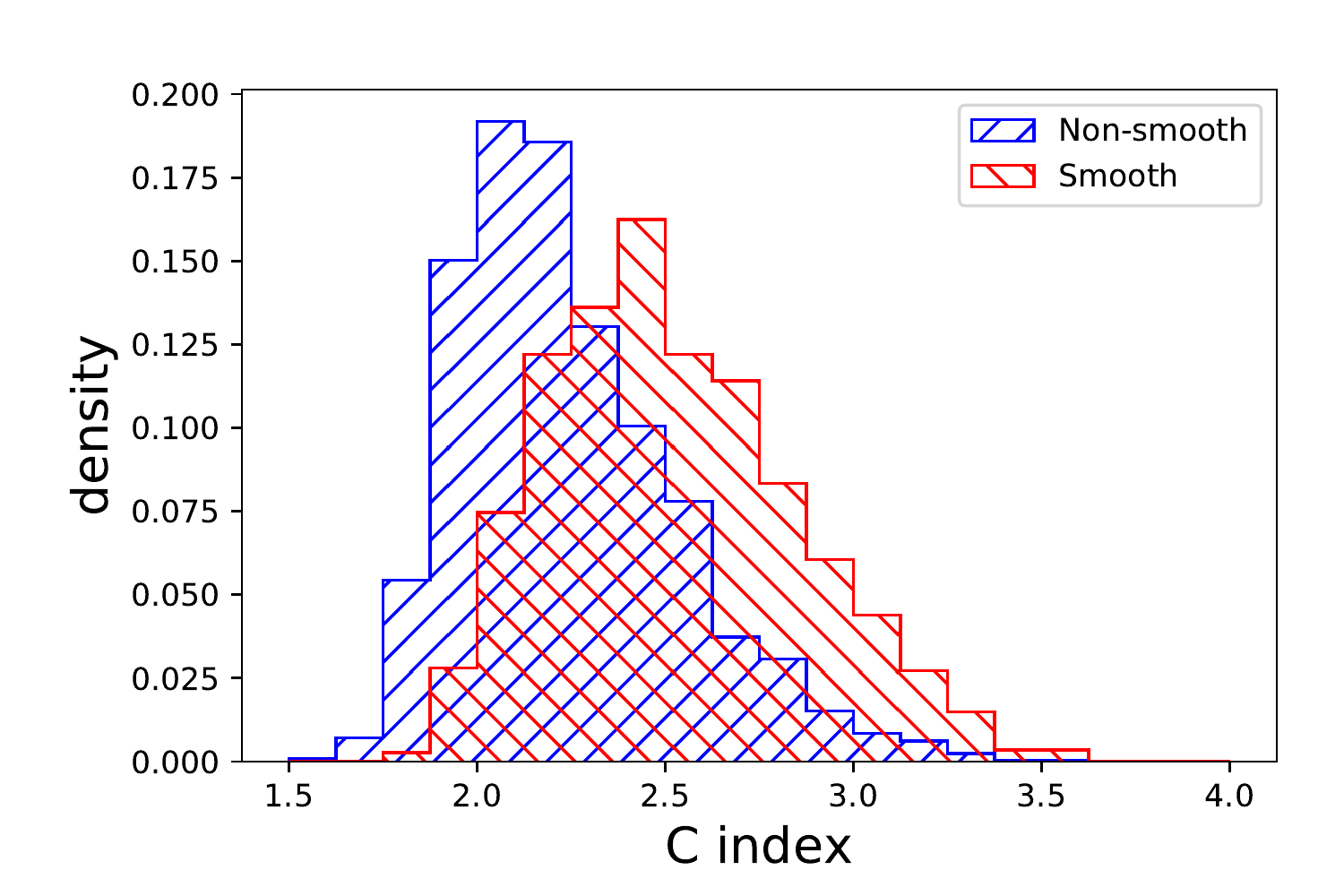}
\caption{Number distribution of non-smooth and smooth galaxies for S\'{e}rsic $n$ (upper) and C-index (lower). We normalized the number distribution by the total number of each population.
\label{fig:histnC}}
\end{figure}

We introduce the \textit{visual smoothness} as a parameter describing the appearance of galaxies. In our analysis, smooth and non-smooth galaxies are determined by the debiased vote fraction of each answer to the first question in GZ2 \citep{Willett2013}, {\it ``Is the galaxy simply smooth and rounded, with no sign of a disk?"} This unique classification distinguishes between \ion{H}{1}-rich and \ion{H}{1}-poor population at fixed $M_*$ and SFR, but what determines the visual smoothness of galaxies? To interpret the visual smoothness, we investigate how the visual smoothness correlates with the other two morphological parameters used in this study (S\'{e}rsic $n$ and C-index; see Figure \ref{fig:histnC}).

\begin{figure}
\gridline{\fig{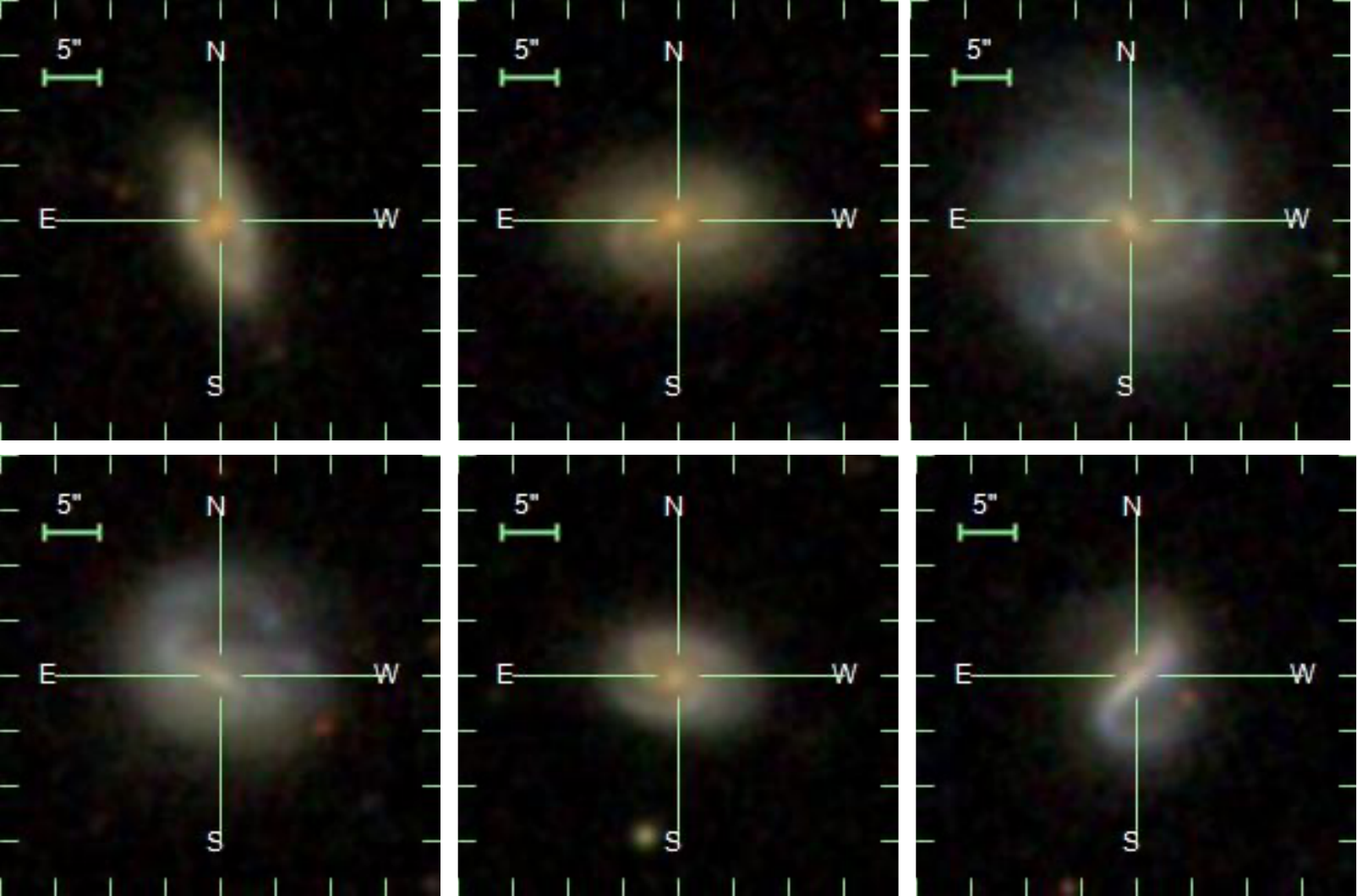}{0.4\textwidth}{Non-smooth}}
\gridline{\fig{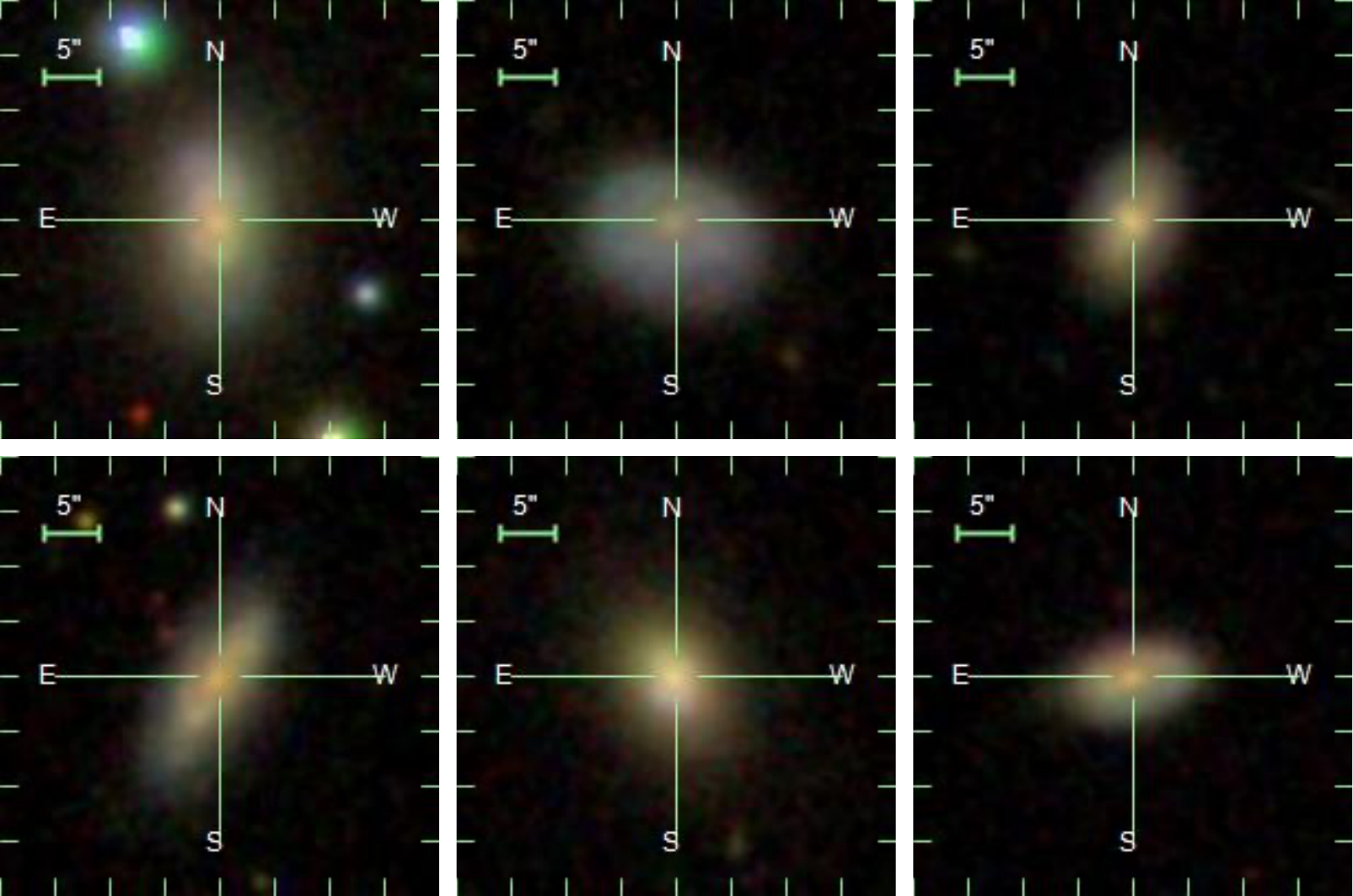}{0.4\textwidth}{Smooth}}
\caption{Examples of optical images of non-smooth (upper 6 panels) and smooth galaxies (lower 6 panels) from SDSS DR12. We note that all these galaxies listed here have fixed stellar mass, \textit{SFR}, S\'{e}rsic $n$, and C-index ($10.0<\log_{10}M_*<10.5$, $0.0<\log_{10}SFR<0.5$, $1.0<n<1.5$, $2.0<C<2.5$).
\label{fig:images}}
\end{figure}

From Figure \ref{fig:histnC}, we realize that both smooth and non-smooth galaxies are distributed over a wide range in S\'{e}rsic $n$ and C-index. The distribution of smooth galaxies has a peak around $n\sim1$, corresponding to a pure exponential disk, and there are non-smooth galaxies at $n\sim4$, which corresponds to the de Vaucouleurs profile. For C-index, smooth galaxies and non-smooth galaxies have a different distribution in the bottom panel of Figure~\ref{fig:histnC}, but the distribution of the smooth galaxies is peaked at $C\sim2.5$, which is actually used to select ``disk'' galaxies in Sec.\ref{subsec:cindex}.
Figure \ref{fig:histnC} suggests that the visual smoothness judged by the citizen scientists is not simply identifying early/late-type or bulge/disk-like morphologies.
We show examples of the optical images of our non-smooth and smooth galaxies from SDSS DR12 in Figure~\ref{fig:images}. The stellar mass, \textit{SFR}, S\'{e}rsic $n$, and C-index of these galaxies are fixed at $10.0<\log_{10}M_*<10.5$, $0.0<\log_{10}SFR<0.5$, $1.0<n<1.5$, and $2.0<C<2.5$. In other words, only visual smoothness is different between the six galaxies in the top and bottom panels of Figure \ref{fig:images}. 

By visually inspecting the optical images in Figure \ref{fig:images}, we realize that many of the smooth galaxies show prominent bulge and disk structure.
It is likely that the citizen scientists' critical features to identify {\it non-smoothness} would be the small structures like spiral arms and/or prominent bar structure within the galaxies.
Indeed, non-smooth galaxies in Figure \ref{fig:images} seem to have spiral arms or prominent bar structure (upper six panels), while smooth galaxies do not have such a small-scale structure (lower six panels).
Although the S\'{e}rsic $n$ and C-index are the morphological indicators commonly used to describe the overall light profile from the core to the outskirts of galaxies, it would not be possible to identify such small-scale structures by those automated parameters. 

Our results suggest that galaxy morphologies defined by S\'{e}rsic $n$ and C-index are not identical to the classification made by human eyes. We note that we are not discussing which is the best indicator of galaxy morphologies. We suggest that the visual smoothness can better distinguish gas-rich and gas-poor populations at fixed stellar mass and \textit{SFR}.
We should also note that our sample is limited to the local universe ($0.01<z<0.05$), so that the physical resolution is high (typically 0.94 kpc at $z=0.033$). This might help citizen scientists to identify small structures in the galaxies.

\begin{figure}
\plotone{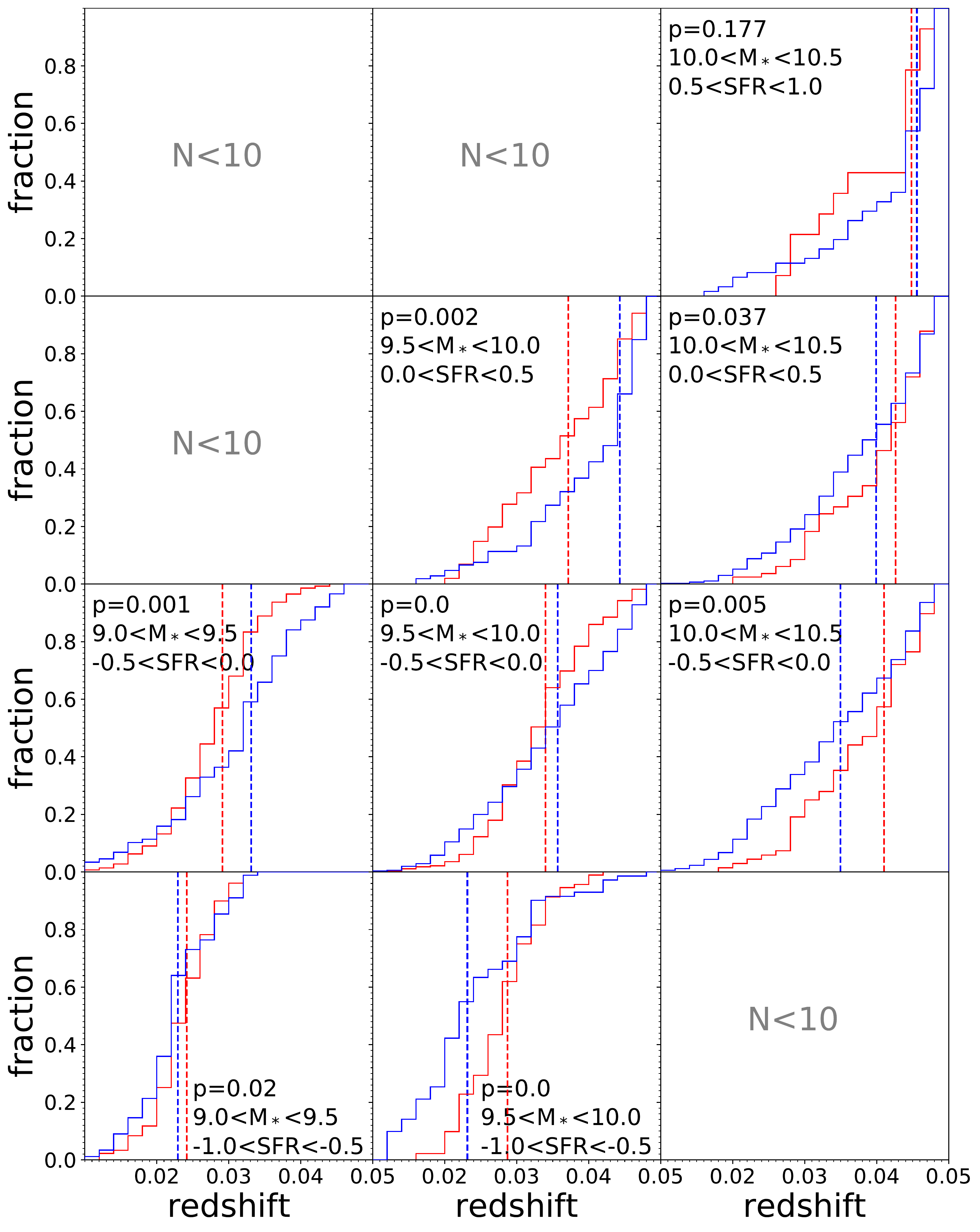}
\caption{The cumulative redshift distribution of our non-smooth (blue) and smooth (red) galaxies for each stellar mass and \textit{SFR} bin. The median value of the redshift in each bin is plotted as dashed vertical lines.
The stellar mass range, \textit{SFR} range (log scale), and the p-value from the KS test are shown in each panel. We note that we require a minimum sample size of 10 for both smooth and non-smooth galaxy subsamples to compute the p-value.
\label{fig:histZ_GZ2}}
\end{figure}

It would also be important to understand the effect (or bias) produced by the redshift of galaxies on the visual classification of the citizen scientists in GZ2. As shown in Figure 4 of \citet{Willett2013}, the fraction of smooth galaxies increases with redshift, while the non-smooth population decreases. This trend is not surprising because the physical resolution becomes poorer for more distant galaxies, and thus it becomes more difficult to identify small-scale structures within more distant galaxies. \citet{Hart2016} corrected the vote fraction by removing the bias produced by the different redshifts. Because we use the debiased vote fraction derived by \citet{Hart2016}, our results would not be affected by the redshift effects. Figure \ref{fig:histZ_GZ2} shows the redshift distributions of our smooth and non-smooth galaxies in each ($M_*$, SFR) bin. This plot suggests that smooth galaxies tend to have higher redshifts than non-smooth galaxies below the star-formation main sequence, while this trend is often reversed for galaxies located above the main sequence. The \ion{H}{1} mass discrepancy between smooth and non-smooth galaxies reported in Figure \ref{fig:comp_morph} (i) is seen regardless of the stellar mass or \textit{SFR} of galaxies, and therefore it is unlikely that the different redshift distribution within each ($M_*$, SFR) bin has significant impacts on the elevated \ion{H}{1} mass excess in non-smooth galaxy population.

Our results suggest that the existence of small-scale structures in the galaxies (e.g., spiral arms) would be the key to determining the \ion{H}{1} gas fraction. 
These small-scale structures might be missed by the C-index and S\`{e}rsic $n$ because they are roughly tracing the overall light profile. In contrast, human eyes are more sensitive to such internal structures, and this would be the main cause of the different morphological dependence of \ion{H}{1} gas mass when we use C-index or visual smoothness for the morphological classification.
Considering the fact that the amount of molecular hydrogen is almost constant \citep{Catinella2018, Koyama2019}, our result suggests that non-smooth galaxies have a larger amount of \ion{H}{1} that is not involved in the star-formation than smooth galaxies, even at the same stellar mass and \textit{SFR}.
More detailed theoretical approaches and/or observations of atomic gas kinematics within the galaxies are necessary to understand the missing link between the small structures likely caused by the local instability (e.g., \citealt{Toomre1977}) and the global properties of \ion{H}{1} gas in galaxies. It is expected that simulations will allow one to trace the motion of gaseous and stellar components within galaxies. However, in most simulations, visual smoothness is not carefully examined because it requires cooperation from many citizen scientists to determine galaxies' visual smoothness. As suggested by \citet{Dickinson2018}, even the most recent cosmological simulations cannot perfectly reproduce morphological properties of present-day galaxies, including the visual smoothness.

\section{Summary}\label{sec:summary}

In this work, we performed \ion{H}{1} spectral stacking analysis for local star-forming galaxies with a wide range in stellar mass, \textit{SFR}, and morphological types. Following the same stacking method presented by \cite{Fabello2011} and \cite{Brown2015}, we show the scaling relation between the stellar mass and the \ion{H}{1} gas fraction (Figure \ref{fig:ALLstack}).
Figure \ref{fig:ALLstack} also suggests that the \ion{H}{1} gas mass depends not only on the stellar mass but also on the star-formation activity of the host galaxies. This result is consistent with \citet{Brown2015}, who concluded that the star-formation activity is a primary driver of the \ion{H}{1} gas content in a galaxy.

We then divide our sample by their morphologies using three morphological parameters; S\'{e}rsic $n$, C-index, and visual smoothness. We separate each subsample into small bins on the SFR--$M_*$ diagram ($\Delta M_*=\Delta SFR=0.5$ dex), and performed the stacking analysis of the radio spectra. Our study revealed that, \textit{at fixed stellar mass and SFR}, \ion{H}{1} gas mass fraction ($F_{{\rm H_I}}$) does not significantly depend on their morphologies when we use C-index for morphological classification, while we do find a significant morphological difference ($\sim$0.7~dex) when we use visual smoothness as a morphological indicator (Figure \ref{fig:comp_morph}). Unfortunately, we could not obtain any clear trend from our analysis with S\`{e}rsic index due to the small number of early-type galaxies.
Because we fixed the stellar mass and the \textit{SFR} of two morphological populations when comparing the \ion{H}{1} gas mass, our result is free from any bias by the stellar mass and the \textit{SFR}. We also investigate the environmental impact on our results. We performed the K-S test and found $p$-values of $>$0.05 for most of the bins. We, therefore, conclude that our results are not affected by galaxies in dense environments like galaxy clusters.

Finally, we study how the visual smoothness correlates with the other two morphological indicators. By comparing the distributions of S\'{e}rsic $n$ and C-index for non-smooth and smooth galaxies, we find that the visual smoothness judged by the citizen scientists is different from the traditional automated distinction between early/late-type or bulge/disk morphologies at least in the local universe (Figure \ref{fig:histnC}). We also compare the optical images of non-smooth and smooth galaxies in Figure \ref{fig:images}. We notice that only non-smooth galaxies have small-scale structures within the galaxies. Therefore we consider that the existence of small-scale structures would be the key to determine the $F_{{\rm H_I}}$ of local galaxies. Because S\'{e}rsic $n$ and C-index are the indicators that describe the overall light profile from the center of the galaxies to the outskirts, the small-scale structure would be missed by those parameters. The potential link between the small-scale structure and the global $F_{{\rm H_I}}$ should be revealed with future spatially-resolved observations and theoretical approaches.

\acknowledgments
We thank the referee for reviewing our paper and providing us with useful comments that improved the paper.

ALFALFA surveys, on which this work is based, were conducted at the Arecibo Observatory. The Arecibo Observatory is operated by SRI International under a cooperative agreement with the National Science Foundation (AST-1100968), and in alliance with Ana G. M\'{e}ndez-Universidad Metropolitana, and the Universities Space Research Association.

Funding for the Sloan Digital Sky Survey (SDSS) and SDSS-II has been provided by the Alfred P. Sloan Foundation, the Participating Institutions, the National Science Foundation, the U.S. Department of Energy, the National Aeronautics and Space Administration, the Japanese Monbukagakusho, and the Max Planck Society, and the Higher Education Funding Council for England. The SDSS Web site is http://www.sdss.org/.

The SDSS is managed by the Astrophysical Research Consortium (ARC) for the Participating Institutions. The Participating Institutions are the American Museum of Natural History, Astrophysical Institute Potsdam, University of Basel, University of Cambridge, Case Western Reserve University, The University of Chicago, Drexel University, Fermilab, the Institute for Advanced Study, the Japan Participation Group, The Johns Hopkins University, the Joint Institute for Nuclear Astrophysics, the Kavli Institute for Particle Astrophysics and Cosmology, the Korean Scientist Group, the Chinese Academy of Sciences (LAMOST), Los Alamos National Laboratory, the Max-Planck-Institute for Astronomy (MPIA), the Max-Planck-Institute for Astrophysics (MPA), New Mexico State University, Ohio State University, University of Pittsburgh, University of Portsmouth, Princeton University, the United States Naval Observatory, and the University of Washington.

This work was financially supported in part by Grants-in-Aid for Scientific Research (No. 18K13588) by the Japanese Ministry of Education, Culture, Sports and Science.


\bibliographystyle{aasjournal}



\end{document}